\documentclass[aip, jap, showpacs, reprint, preprintnumbers,amsmath,amssymb,superscriptaddress,floatfix]{revtex4-2}
\usepackage{graphicx}      % Include figure files
\usepackage{bm}             	% bold math
\usepackage{times}			% Times font
\usepackage{tabularx}
\usepackage{color}
\usepackage{dcolumn}		% Align table columns on decimal point
\usepackage{amsmath}
\usepackage{amssymb}
\usepackage{amsfonts}
\usepackage{times}
\usepackage{tabularx}
\usepackage{xcolor,colortbl}
\usepackage{amsmath}
\usepackage{babel}
\usepackage{multirow}
 \usepackage{ragged2e}
\usepackage[T1]{fontenc}
\usepackage[autostyle]{csquotes}
   \MakeAutoQuote{‘}{’}
\usepackage{csquotes} 

\makeatletter
\newcommand\xleftrightarrow[2][]{%
  \ext@arrow 9999{\longleftrightarrowfill@}{#1}{#2}}
\newcommand\longleftrightarrowfill@{%
  \arrowfill@\leftarrow\relbar\rightarrow}
\makeatother
\begin{document}

\title{\centering{Unidirectionality of spin waves in Synthetic Antiferromagnets}} 
%Dispersion relation and the origin of non-reciprocity in synthetic antiferromagnetic system: Simulation and Measurements  
%Giant nonreciprocity of spin waves in CoFeB/Ru/CoFeB 
%Unidirectional spin waves in CoFeB/Ru/CoFeB Synthetic Antiferromagnet
%Evidence of giant nonreciprocity in symmetrical synthetic antiferromagnet

\author{F. Millo}
\email{florian.millo@universite-paris-saclay.fr}
\author{J.-P. Adam}
\author{C. Chappert}
\author{J.-V. Kim}
\author{A. Mouhoub}
\affiliation{Universit\'e Paris-Saclay, CNRS, C2N, 91120 Palaiseau, France}

\author{A. Solignac}
\affiliation{SPEC, CEA, CNRS, Université Paris-Saclay, 91191 Gif-sur-Yvette, France}
\author{T. Devolder}
\affiliation{Universit\'e Paris-Saclay, CNRS, C2N, 91120 Palaiseau, France}
\date{\today}                                           

\begin{abstract}
We study the frequency non-reciprocity of the spin waves in symmetric CoFeB/Ru/CoFeB synthetic antiferromagnets stacks set in the scissors state by in-plane applied fields. Using a combination of Brillouin Light Scattering and propagating spin wave spectroscopy experiments, we show that the acoustical spin waves in synthetic antiferromagnets possess a unique feature if their wavevector is parallel to the applied field: the frequency non-reciprocity \textcolor{black}{due to layer-to-layer dipolar interactions} can be so large that the acoustical spin waves transfer energy in a unidirectional manner for a wide and bipolar interval of wavevectors. 
Analytical modeling and full micromagnetic calculations are conducted to account for the dispersion relations of the optical and acoustical spin waves for arbitrary field orientations. Our formalism provides a simple and direct method to understand and design devices harnessing \textcolor{black}{unidirectional propagation of} spin waves in synthetic antiferromagnets. 

% \textcolor{red}{Florian: To keep track of changes, please write your name following with the comments/suggestions/critics/corrections - As done in this line. One color choice $=$ one editor.}

%Brillouin Light Scattering measurements are performed in this system were a large frequency non-reciprocity is evidenced. The reason to this large non-reciprocity comes from the layer-to-layer dipolar interaction and further leading to unidirectionality, the one-way flow of energy in the system. Unidirectionality of spin waves were investigated through full micromagnetic calculations and theoretical models.

%We provide a manual to the full micromagnetic calculations of dispersion relations, where two methods are used: one for mode-resolved dispersion relations and second for sign-of-k-resolved dispersion relations. We compare full micromagnetic calculations with theoretical formulas based on the dynamical matrix multilayer calculation and Ishibashi's model. An exhaustive study is made to grasp the agreements and disagreements of the models, calculations and measurements. We provide angle-dependence dispersion relations formulas which allows to predict the behaviour of the system at small fields and low wavevectors that are not present in recent studies.
\end{abstract}

\maketitle
Spin waves (SW), introduced by F. Bloch\cite{bloch_zur_1932}, are the elementary excitations of the magnetic order parameter. They have interesting properties such as non-linear effects, anisotropic propagation and frequency non-reciprocity (NR) \cite{rezende_fundamentals_2020, demokritov_spin_2017, gallardo_spin-wave_2021}. %We focus on nonreciprocal behavior of SWs. 
NR is the situation where the frequency $\omega$ of a SW of wavevector $\vec{k}$ changes upon reversing the direction of phase propagation\cite{nortemann_microscopic_1993}. The asymmetry of the dispersion relation $\omega(\vec{k}) \neq \omega(-\vec{k})$ can be harnessed to design devices with very peculiar behaviors such as magnonic diodes \cite{grassi_slow-wave-based_2020, lan_spin-wave_2015}, directional spin wave emitters~\cite{bracher_creation_2017,osuna_ruiz_unidirectional_2021}, \textcolor{black}{curved spin wave antennas\cite{albisetti_optically_2020}, chiral magnonic logic devices\cite{chen_excitation_2019}} and passive non-reciprocal filters \cite{qin_nanoscale_2021}.

Layer-to-layer dipolar interactions in a multilayer can lead to large NR \cite{stamps_spin_1994}. This requires some contrast of magnetization -- orientation or magnitude -- within the thickness of the multilayer \cite{verba_wide-band_2019}. The contrast is maximal for synthetic antiferromagnets (SAFs), i.e. two ferromagnetic layers separated by a spacer layer that mediates an effective interlayer interaction of coupling parameter $J<0$ favoring an antiparallel state. In SAFs, SWs have two precession modes\cite{stamps_spin_1994} namely acoustical (in-phase) and optical (out-of-phase). Owing to the strong interest on non-reciprocal SWs modes, the analysis of the dispersion relations in SAFs has already been undertaken by several authors \cite{gallardo_spin-wave_2021, verba_wide-band_2019, gallardo_reconfigurable_2019, ishibashi_switchable_2020, shiota_tunable_2020, franco_enhancement_2020, kamimaki_parametric_2020}. 
Analytical expressions were first proposed for a symmetric SAF in refs.~[\onlinecite{verba_wide-band_2019}, \onlinecite{gallardo_reconfigurable_2019}] at zero applied field. Expressions of $\omega(\Vec{k})$ valid at finite fields were then provided by Ishibashi et al. \cite{ishibashi_switchable_2020}. They were successfully used for a qualitative discussion of experimental results. 
\textcolor{black}{One special character of non-reciprocity is unidirectionality, achieved when reversing the sign of the wavevector does \textit{not} reverse the direction of the group velocity $\vec \nabla_{\vec k} (\omega)$, such that the energy carried by a wavepacket is truly "unidirectional": SW can carry energy in one direction but not its opposite. The concept of unidirectional SWs and its potential applications has been reviewed in ref.~[\onlinecite{chen_unidirectional_2022}], which mentions examples of successful implementations on various configurations.\\
In this paper, we show that when a SAF is set in a scissors state, its acoustical spin wave mode can have a very unique dispersion relation: there exists field conditions and wavevector orientations for which unidirectionality is achieved, and in addition is reconfigurable. We demonstrate switchable unidirectionality of acoustical spin waves, in line with previous observations of reconfigurable frequency non-reciprocity in SAF (refs.~[\onlinecite{gallardo_reconfigurable_2019}, \onlinecite{ishibashi_switchable_2020}]). Furthermore, we develop an easy-to-use formalism validated by simulations and measurements. Our results and the associated understanding could then be used to design optimally non-reciprocal devices.}\\
%One still needs to develop an easy-to-use formalism validated by simulations and measurements. The associated understanding could then be used to design optimally non-reciprocal devices.
%In this paper, we show that when a SAF is set in a scissors state, its acoustical spin waves can  
%have a very unique dispersion relation: there exists field conditions and wavevector orientations for which reversing the sign of the wavevector does \textit{not} reverse the direction of the group velocity $\vec \nabla_{\vec k} (\omega)$, such that the energy carried by a wavepacket is truly "unidirectional": SW can carry energy in one direction but not its opposite. 
%\textcolor{black}{The concept of unidirectional SWs and its potential applications has been reviewed in ref.~[\onlinecite{chen_unidirectional_2022}].}\newline
The paper is organized as follows. In section \ref{evidence}, we evidence this unique NR feature by Brillouin Light Scattering (BLS); we then design a device that benefits from this feature. In this device, inductive propagating spin wave spectroscopy (PSWS) experiments demonstrate that the energy flow associated with SWs can be fully unidirectional. 
In section \ref{AnalyticalSection} we derive an approximate description of the group velocities, enabling a physical understanding of the peculiarities of the SW propagation within SAFs. These expressions are compared with ref.~[\onlinecite{ishibashi_switchable_2020}] and with our full micromagnetic calculations in section \ref{MicromagneticsSection}. Finally, their accuracy is discussed.
\section{Evidence of unidirectionality of spin waves in symmetric SAF} \label{evidence} %%
\subsection{Materials: synthetic antiferromagnets} 
We use symmetric SAFs of composition: substrate / Ta(6 nm) / Co$_{40}$Fe$_{40}$B$_{20}$($t_\textrm{mag}$) /  Ru(0.7 nm) / Co$_{40}$Fe$_{40}$B$_{20}$($t_\textrm{mag}$) / Ru(0.4 nm) / Ta(3 nm) where $t_{\textrm{mag}} = 17$ nm. The properties of the SAFs were characterized in refs.~[\onlinecite{mouhoub_exchange_2023}, \onlinecite{ seeger_inducing_2023}], notably to derive %the interlayer exchange energy $J$ through Ru, as well as the magnetization $\mu_{0}M_{s}$, the exchange stiffness $A_{\textnormal{ex}}$, the anisotropy and the Gilbert damping $\alpha$ of the CoFeB layers. We define 
the interlayer exchange field $H_{J} = -\frac{2J}{\mu_{0}  M_{s} t_\textrm{mag}}$, where $M_s$ is the saturation magnetization.\\
Two different substrates were needed: Y-cut LiNbO$_{3}$ substrates, which are optimal for microfabrication\cite{devolder_measuring_2022}, are used for electrical measurements, while naturally oxidized silicon substrates are optimal for Brillouin Light Scattering experiments. The corresponding samples have slightly different properties. Both samples have in-plane uniaxial anisotropies much smaller than their respective spin-flop fields. They share the same saturation magnetization $\mu_{0}M_{s}$ = 1.71 T and the same damping $\alpha \approx 0.0045$, although different interlayer exchange fields: $\mu_{0} H_{J} = 148$ mT for Y-cut LiNbO$_{3}$ and $\mu_{0} H_{J} = 78$ mT for oxidized silicon substrates (more details are given in appendix). Since we will focus on SWs of long wavelengths where the shapes of the dispersion relations are mainly determined \textcolor{black}{by layer-to-layer} dipolar effects\cite{nortemann_microscopic_1993, gallardo_spin-wave_2021, gallardo_reconfigurable_2019, ishibashi_switchable_2020, henry_propagating_2016}, we will consider that the two samples are essentially comparable.

\subsection{Spin wave dispersion relations} %%%%%%%%%%%%%

We measured the dispersion relations of the SWs within the SAF by performing wavevector-resolved BLS spectroscopy. The experimental configuration is sketched in Fig.~\ref{FigureBLS}(b). The applied field $\mu_0 |\vec{H}_{0}|=$ 25 mT, higher than the spin-flop field ($\mu_0 |\vec{H}_{\textrm{SF}}|=$ 8 mT), was chosen to set the SAF in one of the two degenerate scissors states\cite{ishibashi_switchable_2020} sketched in Fig.~\ref{FigToggling}(c). The BLS laser beam impinges the sample surface with an incidence $\theta \in [5^{o}-30^{o} ]$ with respect to the normal and we analyse the back-scattered photons. The scattering conserves the total momentum, so the wavenumber of the magnons annihilated (anti-Stokes process) or created (Stokes process) in the scattering process obeys ${k}_ {x}= \pm \frac{4 \pi}{\lambda} \sin\theta \in [\pm 2, \pm 12] \; \textrm{rad/}\mu\textrm{m}$ with a frequency shift allowing to construct $\omega(k)$. 
Fig.~\ref{FigureBLS}(c) reports the measured dispersion relations $\omega(\Vec{k}) / 2 \pi $ for $\Vec{k}_x \; || \; \Vec{H}_{0}$. Two points are worth emphasizing:

\begin{figure}
\includegraphics[scale=0.22]{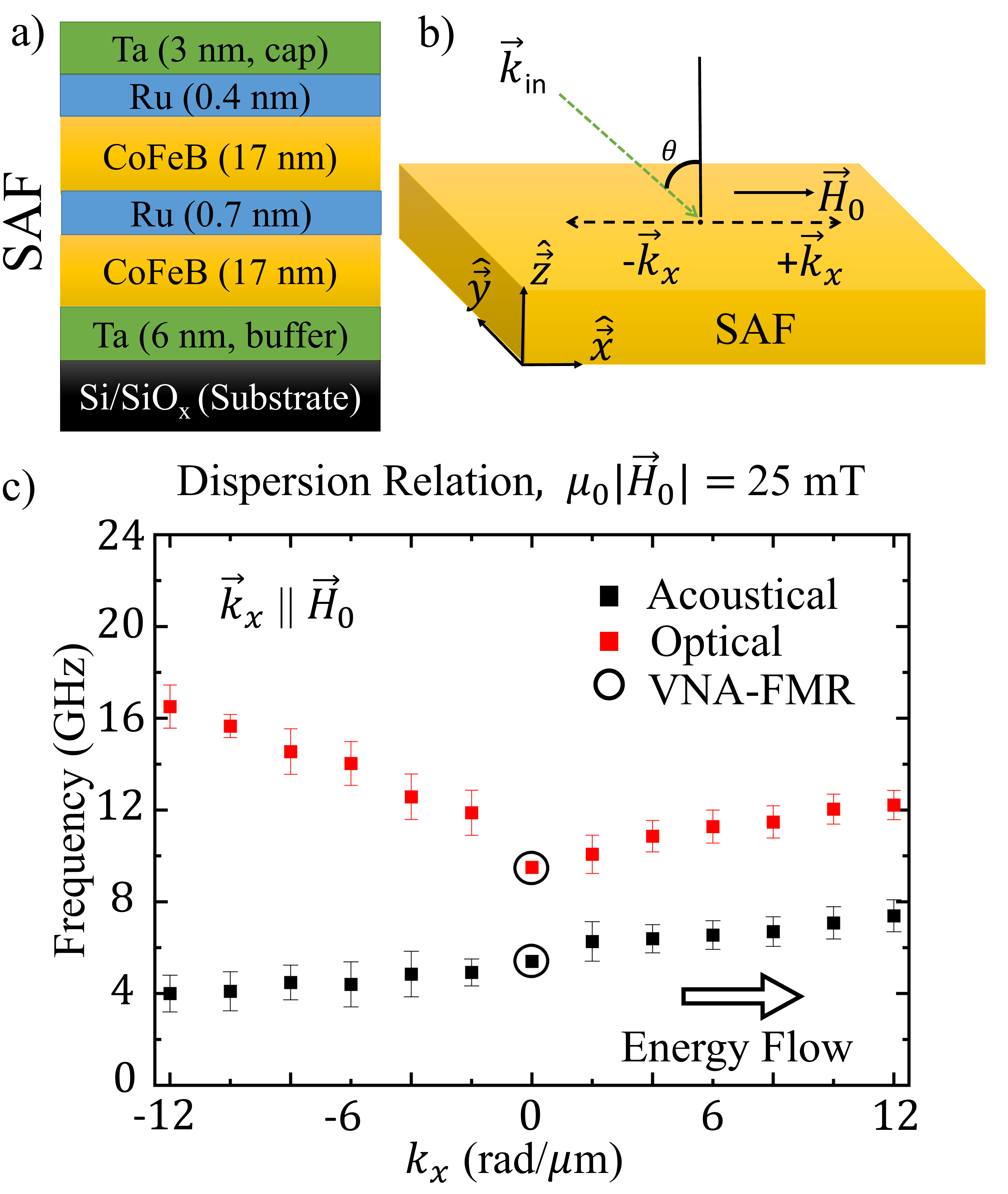}
\caption{a) Multilayer stack of the synthetic antiferromagnet used for the BLS experiments. b) BLS geometry: applied field $\Vec{H}_{0}$ parallel to the $\hat{\vec{x}}$ axis, plane of incidence (xz), wavelength of laser beam $\lambda=532$ nm. c) Dispersion relations are measured by BLS for $\vec{k}_{x} \parallel \vec{H}_{0}$. The uniform resonances (circles at $k$=0) were measured by VNA-FMR. The error bars are the linewidths (full width at half maximum). The black arrow emphasizes that the group velocity of the acoustical SWs always points toward the positive side, irrespective of the sign of the wavevector. %V_{g} = 1.25 km/s
}
\label{FigureBLS}
\end{figure}

% importand features, to be noticed in disp relations
\begin{itemize}
    \item The optical spin wave mode has a tilted (i.e. non-symmetric) $\vee $-shaped dispersion relation [see Fig.~\ref{FigureBLS}(c)]: it is non-reciprocal. The NR is classical in the sense that the group velocity has always the same sign than the wavevector, i.e. it is a "forward" wave, in agreement with the forthcoming modeling (Table.~\ref{TableFormulas} and section \ref{MicromagneticsSection}).    
    \item In contrast, the dispersion relation of the acoustical spin wave branch is quasi-linear in the range explored here [-12,12] rad/$\mu$m, a fact that is very unusual for a wave. The group velocity of these SWs is always positive, whatever the sign of the wavevector. This is a unique feature of the acoustical SWs of a SAF set in the scissors state. As a result, the energy carried by a wave-packet of acoustical SWs is expected to propagate in a one-way manner. 
\end{itemize}

It is also interesting to note that the frequency NR $\delta f = \frac{1}{2 \pi}(\omega(|\Vec{k}|) - \omega(-|\Vec{k}|))$ of the two modes are of comparable magnitudes although with opposite signs. For $k = 2$ rad/$\mu$m, the NRs amount already to $\delta f_{ac} = 1.35$ GHz and $\delta f_{op} = -1.81$ GHz. For $k = 12$ rad/$\mu$m, the NRs are as large as $\delta f_{ac} = 3.39$ GHz and $\delta f_{op} = -4.29$ GHz. These very large values demonstrate that the SAF has an exceptional NR.\\
\textcolor{black}{Gallardo et al.\cite{gallardo_reconfigurable_2019} predicted by theoretical and micromagnetic simulations and measured by BLS a high NR in frequency in a bilayer systems due to the layer-to-layer dipolar interactions. They found that this high NR of acoustical SWs can be controlled by equilibrium configuration and by the geometry of the system. Our results fully agree with their study.} For the acoustical mode, the NR even achieves unidirectionality of the energy flow [see Fig.~\ref{FigureBLS}(c)]. Let us illustrate this point using electrical measurements. 

\subsection{Energy flow carried by propagating spin waves}
\begin{figure}
\includegraphics[width=8cm]{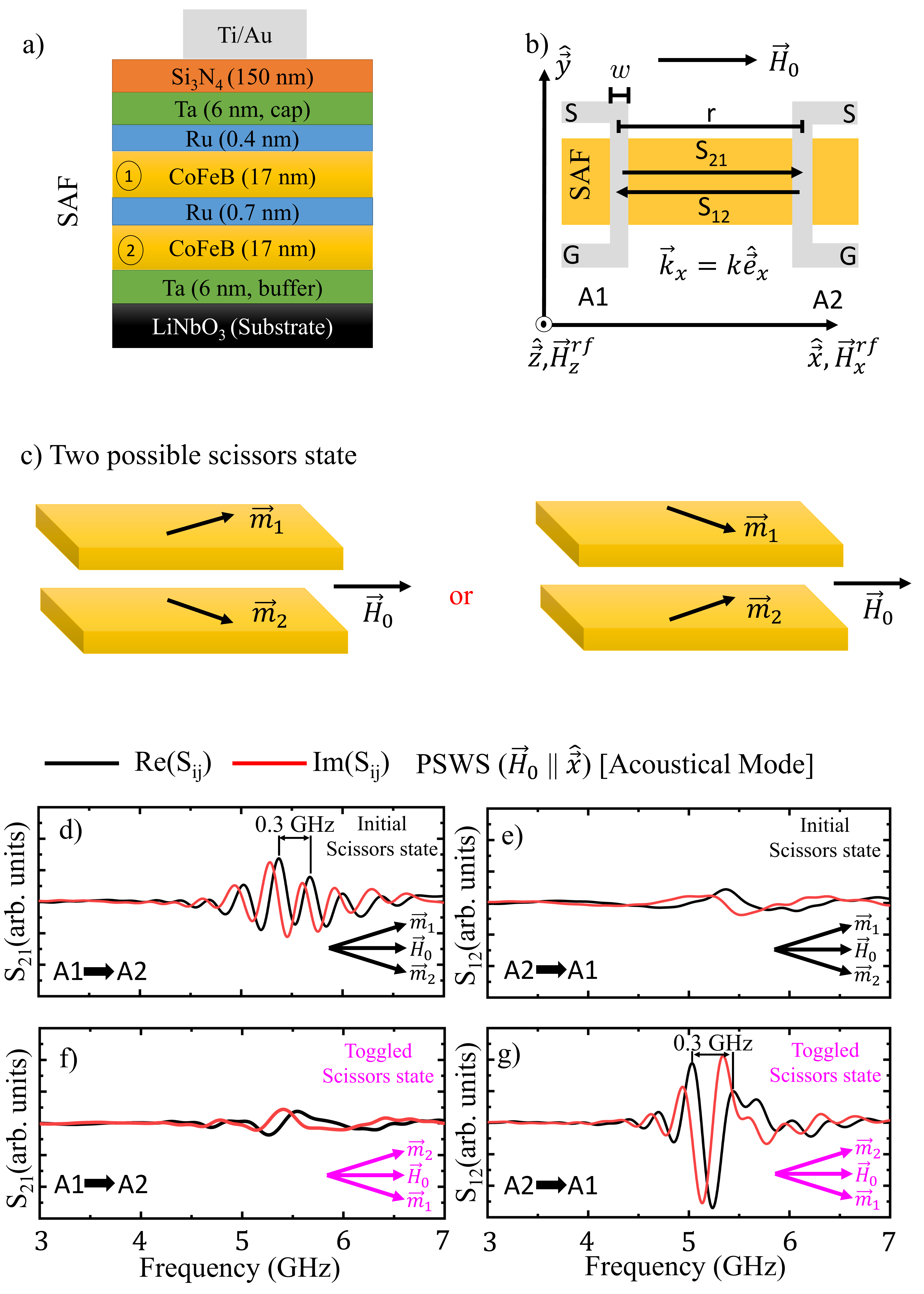}
\caption{ a) Stack of synthetic antiferromagnetic film patterned into devices for propagating spin wave spectroscopy experiments. b) Geometry of the device. Gray: single-wire antennas with $r = 6\;\mu\textrm{m}\;\textrm{and}\; w = 1.8 \;\mu\textrm{m}$. Antenna 1 (A1) and Antenna 2 (A2) are connected to a VNA to collect transmission [$S_{12}, S_{21}$] parameters. c) Two possible scissors state of the system. d) and e) Forward and backward transmission parameter measured for wave propagating from A1-to-A2, A2-to-A1 respectively and at one of two possible degenerate scissors state. 
f) and g) Forward and backward transmission parameter measured for propagating SWs after toggling the scissors state.  
%\textcolor{black}{$V_{g} = |\Delta f| \times r = 1.9$  km/s}.
}
\label{FigToggling}
\end{figure}

In this section, we perform propagating spin wave spectroscopy measurements to identify in which direction the spin waves effectively transfer energy. The measurement relies on the device sketched in Fig.~\ref{FigToggling}(a) and (b). The device is a SW conduit made of a $20~\mu$m-wide SAF stripe covered by a thick insulating layer and two inductive antennas. These rf antennas have a width of $w= 1.8 ~\mu$m and a center-to-center distance of 6 $\mu$m in the $\hat {\vec x}$ direction. The antennas can be considered as infinite in the $\hat {\vec y}$ direction and therefore they only couple to SWs with wavevectors directed in the $\hat {\vec x}$ direction, as seen in Fig.~\ref{FigToggling}(b). When connected to a vector network analyzer, the antennas emit a circulating rf field $\Vec{H}^{rf}(x,z)$ with components along both $\hat{\vec{x}}$ and $\hat{\vec{z}}$ axes. Thanks to these two components the rf field can couple with the acoustical spin waves even when $\Vec{H}_{0} \parallel \hat{\vec{x}}$. The applied field $\mu_{0} |\Vec{H}_{0}| = 41$ mT was chosen to lead to the same uniform resonance acoustical frequency $f^{k=0}_{ac} = 5.4$ GHz as in Fig.~\ref{FigureBLS}(c).

The 4 measurements of Fig.~\ref{FigToggling}(d, e, f, g) are done for exactly the same applied field $\Vec{H}_{0} \parallel \hat{\vec{x}}$. This field sets the SAF in one of two degenerate scissors state [see Fig.~\ref{FigToggling}(c)] where the equilibrium magnetizations of each layer [see Fig.~\ref{FigToggling}(c)], $\vec{m}_1$ and $\vec{m}_2$, scissors toward the direction of the applied field $\vec{H_{0}}$. The scissors state that is effectively obtained is unknown: $(\vec{m}_1 \times \vec{m}_2). \hat{\vec{z}}$ can be either positive or negative. However, a toggle switching experiment\cite{savtchenko_method_2003} can switch the scissors from one state to the other. 

The experiment is done in four steps. First we apply $\vec{H}_{0}$ in the +$\hat{\vec{x}}$ direction and characterize the sample [see Fig.~\ref{FigToggling}(d, e)]. The SAF is in one of the two possible scissors states [see Fig.~\ref{FigToggling}(c)]. Then, we rotate $\vec{H}_{0}$ by 180$^{o}$ degrees. The scissors state follows the rotating field. We then take back $\vec{H}_{0}$ to 0 mT, then apply again $\vec{H}_{0}$ in the +$\hat{\vec{x}}$ direction: the newly obtained "toggled" scissors state [see Fig.~\ref{FigToggling}(f,g)] is the inverse of the initial one. Finally, we perform the characterization exactly as for the initial state.

Fig.~\ref{FigToggling}(d) shows that in the first scissors state, the forward transmission parameter is oscillatory and strong, while in the toggled scissors state [Fig.~\ref{FigToggling}(f)] it is weak and non-oscillatory, with a quasi-Lorentzian shape. The oscillatory character is indicative\cite{devolder_measuring_2021} that the forward transmission arises from propagating waves. The quasi-Lorentzian shape is indicative \cite{sushruth_electrical_2020} that it arises from quasi-uniform SWs excited under the receiving antenna by the rf field of the exciting antenna, \textcolor{black}{cf. Fig.~\ref{FigToggling}(e) and Fig.~\ref{FigToggling}(f)}, without significant propagation of the involved SWs.

This interpretation -- propagating SWs in the sole forward direction -- is corroborated by the frequency contents of the transmission signals. Indeed the backward signal is finite only near the uniform resonance $f_{ac}^{k=0}=5.4$ GHz within its linewidth. In contrast, the forward transmission parameter spans over frequencies both much below and much above $f_{ac}^{k=0}$. This stems from the linear character of the acoustical dispersion relation: the SWs with $k<0$ have frequencies below $f_{ac}^{k=0}$ and the SWs with $k>0$ are above $f_{ac}^{k=0}$. They have both a group velocity pointing towards the receiving antenna, such that this leads to a one-way transfer of energy between the antennas, in the same direction for frequencies above and below $f_{ac}^{k=0}$.

Interestingly also, when toggling the SAF from one of degenerate scissors state to the other, we observe that the direction in which the acoustical spin waves can transfer energy is reversed [compare the panels of Fig.~\ref{FigToggling}(d,f) and Fig.~\ref{FigToggling}(e,g)]. As we shall see later, this arises from the fact that the dispersion relation of acoustical spin waves undergoes the transformation $\vec{k}_{x} \rightarrow -\vec{k}_{x}$ when the two magnetizations of the SAF are toggled while keeping $\Vec{H}_{0}$ applied along $+\hat{\vec{x}}$ direction. This experiment can be interpreted as switching the unidirectionality of the acoustical SWs, to be compared to the experiment of ref.~[\onlinecite{ishibashi_switchable_2020}] that switches the non-reciprocity of optical SWs.

\section{Approximate group velocities}  \label{AnalyticalSection} %%%%%
Let us model the dispersion relations of SW in SAF and their dependence over the field orientation. We shall make approximations to get insightful expressions.
\begin{table*} %%%%%%%%%%%%%%%%%%%%%%%%%%%%%%%%%%%%%%%%%%%%%%%%%%%%%%%%%%%%%%%%%%%
\caption{Group velocities of SWs in a SAF for a  2-macrospin ground state in the long wavelength limit. The SAF is isotropic and fully symmetrical. The field $\vec{H}_0$ is applied along the $\hat{\vec{x}}$ direction. The group velocities are meant as $\frac{d\omega}{dk}$, irrespective of the sign of $k$. They are given in units of $\frac{1}{2} \gamma_0 M_s t_{\textrm{mag}}$. The colored cells highlight the situations when the SWs show frequency non-reciprocity.  The $\dagger$ symbol recalls that the Taylor expansion to first order in $kt_{\textrm{mag}}$ is not legitimate when the acoustical eigenfrequency vanishes. Note that the sign of non-reciprocal terms within the group velocities are reversed when toggling the magnetizations of the two layers of the SAF.}
\label{VgSAF}
\begin{tabular}{| c | c | c | c | c | c |}
 \hline
Field  & $H_0 = 0$  & $0 < H_0< H_j$ &  $H_0 \geq H_j$   \\ \hline
State  & Antiparallel  & Scissors & Parallel    \\ \hline
\noalign{\vskip 2mm} \hline
\textbf{Acoustical branch} \\ \hline

$v_g^\textrm{ac}$, $\vec{H}_0 \parallel \vec{k}_{x}$, %$\phi=\frac{\pi}{2}$ 
$k_{x}>0$ & \cellcolor{blue!15}  $-1$ $\dagger$  &\cellcolor{blue!15} $-  \left(
\sqrt{1 - \left(\frac {H_0} {H_j}\right)^2} + \frac{H_0} {\sqrt{H_j (M_s+H_j)}}
\right)$ &   $ - \sqrt{\frac{H_0}{H_0+M_s}}  $ ~($\Lambda-$shaped)\\ \hline

$v_g^\textrm{ac}$, $\vec{H}_0 \parallel \vec{k}_{x}$, %$\phi=\frac{\pi}{2}$ , 
$k_{x}<0$& \cellcolor{blue!15} {$-1$} $\dagger$  & \cellcolor{blue!15} {$-  \left(\sqrt{1 - \left(\frac {H_0} {H_j}\right)^2} - \frac{H_0} {\sqrt{H_j (M_s+H_j)}}
\right)$} & {$ + \sqrt{\frac{H_0}{H_0+M_s}} $} ~($\Lambda-$shaped) \\ \hline

$v_g^\textrm{ac}$, $\vec{H}_0 \perp \vec{k}_{x}$, %$\phi=0$
$\forall k_{x}$ &  0 & 
$ \textrm{sgn($k_{x}$)} \frac{H_0}{H_j} \frac{M_s}{\sqrt{H_j(H_j+M_s)}}$ &  $ \textrm{sgn($k_{x}$)}  {\frac{M_s}{\sqrt{H_0(H_0+M_s)}}}$   \\  
\noalign{\vskip -0.11mm}
  & & ($\vee-$ shaped dispersion) & ($\vee-$ shaped dispersion)  \\  \hline

$v_g^\textrm{ac}$, $\{\vec{k}_{x}, \vec{H}_0\}= \varphi$, %$\phi=ANY$
$k_{x}>0$ & \cellcolor{blue!15} $- \cos \varphi $ $\dagger$ & \cellcolor{blue!15} $ v_g^{ac} \bigr|_{H_0=0}  - \frac{H_0}{H_j} \left( \frac{ H_j - (M_s+H_j) \sin^2 \varphi}{\sqrt{H_j(M_s+ H_j)}}  \right) + O(H_x^2)$  
&    $ \left(\frac{\sqrt{M_s+H_0}} 
   {\sqrt{H_0}}\sin^2 \varphi-\frac{\sqrt{H_0}}{\sqrt{M_s+H_0}}\right)$  \\ \hline

\noalign{\vskip 1.5mm}
\hline

\textbf{Optical branch} \\ \hline

$v_g^\textrm{op}$, $\vec{H}_0 \parallel \vec{k}_{x}$, %$\phi=\frac{\pi}{2}$ 
$k_{x}>0$ & \cellcolor{blue!15} $+ \left(\sqrt {\frac{M_s} {H_j}}+1\right)$  & \cellcolor{blue!15} $+ \left(\sqrt \frac{{M_s}}{{H_j}}+1\right) \sqrt{1 - \left(\frac {H_0} {H_j}\right)^2}$  &  0    \\ \hline

 $v_g^\textrm{op}$, $\vec{H}_0 \parallel \vec{k}_{x}$, %$\phi=\frac{\pi}{2}$, 
$k_{x}<0$ & \cellcolor{blue!15}{$- \left(\sqrt {\frac{M_s} {H_j}}-1\right)$}   & \cellcolor{blue!15} {$- \left(\sqrt \frac{{M_s}}{{H_j}}-1\right) \sqrt{1 - \left(\frac {H_0} {H_j}\right)^2}$}     &  {0}   \\ \hline
 
 $v_g^\textrm{op}$, $\vec{H}_0 \perp \vec{k}_{x}$, %$\phi=0$, 
 $\forall k_{x}$ & {0} 
&
{0} &  {0} \\  
\noalign{\vskip -0.5 mm}
 (flat dispersion) & & &  \\  \hline
$v_g^\textrm{op}$, $\{\vec{k}_{x}, \vec{H}_0\}= \varphi $, $k_{x}>0$ & \cellcolor{blue!15} $ + \left(\sqrt {\frac{M_s} {H_j}}\cos\varphi+1\right) \cos\varphi$ & \cellcolor{blue!15} $ v_g^{op} \bigr|_{H_0=0} + O(H_0^2)$  & 0 \\ \hline

\noalign{\vskip 2mm}

\end{tabular}
\label{TableFormulas}
\end{table*}
\subsection{Simplifying assumptions and methods}
We describe the ground state in the 2-macrospin approximation, i.e. we assume that the SAF comprises two films (labelled 1 and 2 [see Fig.~\ref{FigToggling}(a)]) that are uniformly magnetized across their thickness. %The ground state is $\vec M_\textrm{dc}=\{\vec { M}_1^\textrm{dc}, \vec M_2^\textrm{dc} \}$. 
For an applied field $|H_{0}| < H_J$, the SAF is in one of the two degenerate scissors states \cite{stamps_spin_1994, devolder_spin_2012}. Note that to consider the other scissors state [obtained by swapping of the magnetizations of the top (t) and bottom layers (b)], one just needs to rotate the sample by 180$^{o}$ around the $\hat{\vec{y}}$ axis, which changes $+\hat{\vec{x}}$ into $-\hat{\vec{x}}$ and $+\vec{k}_{x}$ into $-\vec{k}_{x}$. The measurement geometry does not change in this process, but the top layer become the bottom one and the antennas 1 and 2 get exchanged. The measured dispersion relations have thus to obey the bottom-top swapping property: \begin{equation}\omega_{tb}(\vec k) = \omega_{bt}(- \vec k),\end{equation} 
as confirmed from the PSWS characterizations [see Fig.~\ref{FigToggling}(d,e,f,g)].
The dynamical matrix of the system can be calculated following the standard methodology \cite{%grunberg_magnetostatic_1981, 
nortemann_microscopic_1993, zhang_angular_1994, grimsditch_magnetic_2004, giovannini_spin_2004} and must be augmented with the contribution of finite wavevectors to the self- and mutual demagnetizing effects of the two magnetic layers (see Eq. 25 and 26 of ref.~[\onlinecite{henry_propagating_2016}] for a general description, or ref.~[\onlinecite{ishibashi_switchable_2020}] for the specific case of a SAF set in scissors state).
The two eigenvalues of the dynamical matrix are the frequencies of the SWs: the acoustical \cite{stamps_spin_1994} SW mode for which the scissoring state undergoes essentially a rigid rocking, and the optical SW mode for which the scissoring angle breathes. At $k=0$ the eigenfrequencies reduce to: 
\begin{equation}
 \frac {\omega_\textrm{ac}^{k=0}}{\gamma_0}=   H_{0} \sqrt{\frac{M_s+H_{J}}{H_{J}}} \textrm{~and~} \frac{ \omega_\textrm{op}^{k=0}}{\gamma_0} =  \sqrt{\frac{M_s}{H_{J}}} \sqrt{H_{J}^2-H_{0}^2}
 \label{FMRfrequencies}
\end{equation}
where $\gamma_0$ is the gyromagnetic ratio.

Being solutions of a biquadratic equation, the expressions of the frequencies of the optical and acoustical spin waves are heavy (see for instance Eq. 1 and 2 ref. [\onlinecite {ishibashi_switchable_2020}]).  To get linear in $k t_{\textrm{mag}}$  (simple to apply) formulas, these frequencies were Taylor-expanded to first order in $k t_{\textrm{mag}}$ near $kt_{\textrm{mag}}=0$. The noticeable expressions of the group velocities are gathered in Table~\ref{VgSAF}, given in $\frac{1}{2} \gamma_0 M_s t_{\textrm{mag}}$ units. We must again emphasize that we are in a dipolar dominated system where at low $\vec{k}$ , $\lambda/t_{\textnormal{mag}} \gg 1$.
For $ \vec{H}_{0} \perp \vec{k}_{x}$ and $ \vec{H}_{0} \parallel \vec{k}_{x}$, the group velocities have simple expressions that are consistent with previous reports \cite{devolder_measuring_2022, ishibashi_switchable_2020}. 
For wavevectors directed in other directions, getting simple formulas requires a further Taylor expansion assuming $H_0 \ll H_j$ to provide an analytical angular dependence of the $v_g$'s valid at low fields.

\subsection{Peculiarities of the group velocities of a SAF in the scissors state} \label{PeculiaritiesDispersionRelations}
Several points are worth noticing in Table~\ref{VgSAF}. 

%natural scales
(i) At low applied fields, the group velocities of the acoustical SWs follow a natural scale which is $\frac{1}{2} \gamma_0 M_s t_{\textrm{mag}}$. It amounts to 2.5 km/s for our material parameters. This velocity scale is independent from the exchange coupling $J$.\\
In contrast, the group velocities of the optical SWs comprise an additional dimensionless accelerating factor that depends on $J$. For $\vec k \parallel \vec{H}_{0}$ this accelerating factor is either $+\left(\sqrt {\frac{M_s} {H_j}}+ 1 \right)$ or $-\left(\sqrt {\frac{M_s} {H_j}}- 1 \right)$. These are respectively 5.6 and -3.6 for our samples.  %\textcolor{black}{5.68 and 3.68}. 
This is in agreement with BLS study that concluded that optical SWs are faster than acoustical ones when $\vec k \parallel \vec{H}_{0}$ [see Fig.~\ref{FigureBLS}]. \\
\indent(ii) In the scissors state the SWs in a SAF are \textit{almost always} non-reciprocal. The only reciprocal case is when $ \vec{H}_{0} \perp \vec{k}_{x}$. The dispersion is then $\vee$-shaped for the acoustical mode and flat for the optical mode. In the other cases, despite its non-reciprocity, the dispersion relation of the optical SWs is still conventional in the sense that the sign of group velocity and the sign of $k$ are always correlated. For most\footnote{The optical wave has a backward character only in the narrow range of orientations when $\frac{\pi} 2 < \varphi < \cos^{-1} (-\sqrt{\frac{H_j}{M_s}})$. In this case the group velocity is very close to vanish and propagation is not foreseen.} wavevector orientations, the dispersion relation of the optical SWs has the shape of a tilted $\vee$: the group velocity and the wavevector are of same sign and a SW transmits energy in the direction of its wavevector. \\
%conterintuitive propagation vg // k acoustical
\indent(iii) As anticipated from the experimental results, the most counter-intuitive situation is that of acoustical SWs of wavevectors $ \vec{H}_{0} \parallel \vec{k}_{x}$ in the scissors state. In this case, the dispersion relation is quasi-linear across $k=0$, such that the group velocity points towards the same direction \textit{regardless} of the sign of $k$. Acoustical spin waves with $ \vec{H}_{0} \parallel \vec{k}_{x}$ can transfer energy only in one direction. Since toggling the SAF to the other scissor state is equivalent to changing the sign of $k$, this also explains why the direction in which SWs can transfer energy is reversed when toggling the SAF.\\  
\indent(iv) The non-reciprocities of the acoustical and optical modes defined as $\delta f_{ac, op}= f_{ac, op}(k) - f_{ac, op}(-k)$ are predicted to be equal in magnitude but opposite. \\
From Taylor-expanded expressions of the SW frequencies versus $\vec{k}_{x}$ we find:
\begin{equation}
\delta f_{op},~~- \delta f_{ac}=   \gamma_0 M_s k \, t_{\textrm{mag}}\; \sqrt{1 - \left(\frac {H_0} {H_j}\right)^2} \; \cos \varphi
\label{AnalyticalNR}
\end{equation}
This expression is exact for $\phi=0$ and $\pi/2$ but defined at $O(H^2_0)$ otherwise.

(v) Finally we emphasize that the group velocity $\vec \nabla_{\vec k} (\omega)$ is in general not colinear with the wavevector, such that wavepackets with a 2D content of wavevectors will propagate with potentially non trivial deformations like self-focusing and caustics, as already mentioned in ref. [\onlinecite{gallardo_spin-wave_2021}].\\
\indent The expressions gathered in Table \ref{VgSAF} are convenient for the physical understanding. They must however be considered with care: indeed the linear Taylor expansion over $kt_{\textrm{mag}}$ leads to negative (unphysical) frequencies when used at low fields. Besides, the expressions in Table~\ref{VgSAF} suffer from a fundamental limitation: the 2-macrospin ground state is inaccurate for SAFs thicker than the exchange lengths \cite{mouhoub_exchange_2023}. In this case, a gradient of the magnetization orientation can develop within each layer\cite{mouhoub_exchange_2023}, which renders incorrect the eigenfrequencies at $k=0$. The micromagnetic configuration and its SWs must then be calculated numerically using a full micromagnetic framework.
\section{Micromagnetic Simulations of the dispersion relations} \label{MicromagneticsSection}
In this section we will develop micromagnetic simulations based on the Mumax3 software \cite{vansteenkiste_design_2014} to provide more reliable dispersion relations of Si / SiO$_{x}$ / SAF system. 
After describing the geometry, we detail how to selectively excite the acoustic or optic modes. Then we describe a method that allows to differentiate $+k$ (forward phase velocity) and $-k$ (backward) in any non-reciprocal situation. 

\subsection{Simulation geometry}
To mimic experiments, we first define a SAF slab of in-plane dimensions \{$l_{x} \times l_{y} $\} where $l_{x}=l_{y}=4 \pi \mu$m $\approx 12.6~\mu$m and total thickness of 34 nm. 
The slab is replicated 10 times in $\hat{\vec{x}}$ and $\hat{\vec{y}}$ directions to approximate an infinite thin film. To optimize the calculation time, each slab is meshed into $512\times64\times 16$ cells. The slices along the thickness are numbered $i=1,...,16$ upward. An interlayer exchange coupling is applied between slices $i=8 $ and $i=9$ to account for the Ru layer whose thickness is neglected. We first apply an in-plane static field $H_{0}$ and let the system relax to its ground state. As in the experiments, the angle $\varphi$ is defined between the SW wavevector ($k_x$ conventionnaly along $x$), and the field $H_{0}$ %=\{\vec{k}_{x},
[see Fig.~\ref{FigureMicromagneticMethod}(c)].

\subsection{Methods} %%%%%%%%%%
To identify the dispersion relations, we excite the sample with stimuli that bring it out of equilibrium and we analyze the spectral content of the magnetization response.
\subsubsection{Mode-resolved dispersion relations} \label{modeResolvedSection} %%%%%%%%
\begin{figure}
\includegraphics[scale=0.12]{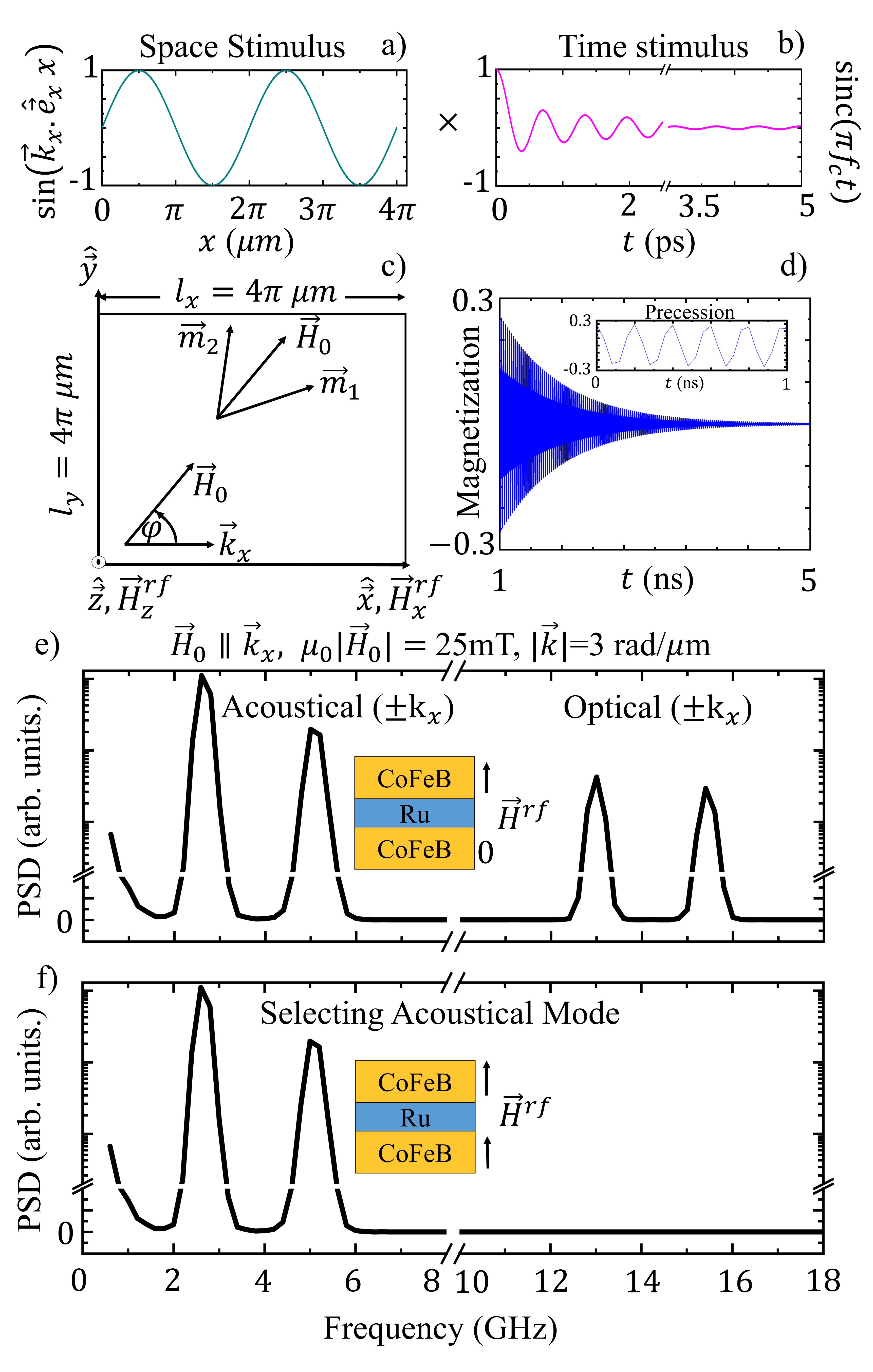}
\caption{Method used to construct the mode-resolved dispersion relations as described in section \ref{modeResolvedSection}.
(a) Spatial dependence and (b) time dependence of the pulsed RF field [Eq.~\ref{simulus1}]. (c) Top view of the simulated system with lateral dimensions of $(4 \pi \times 4 \pi) ~\mu \textrm{m}^2$. The SAF is in the scissors state. The angle between SW wavevector $\vec{k}$ and the DC field $\vec{H}_0$ is $\varphi$. (d) Sketched time-resolved response of the magnetization to the stimulus of (a) for $q(z)=\Theta(i-8.5)$. (e) Power spectral density (PSD) of this response after apodization (Eq.~\ref{fourier}). The maxima correspond to the acoustical and optical modes sharing the common spatial periodicity defined in panel (a). (f): PSDs obtained after the selective excitation of acoustical mode using a uniform $q(z)$. Insets in (e) and (f): sketches of the thickness profiles of the stimuli.}

\label{FigureMicromagneticMethod}
\end{figure}

A frequently implemented stimulus is a short field pulse that imprints a spatial periodicity and contains a broadband frequency content. We choose to apply a pulsed RF field of the form:
\begin{equation}
   \vec{H}^{rf}(x, t, z) = h_{0} \, f(x)\, g(t)\, q(z)\, \hat{\vec{e}}_{z}
    \label{simulus1}
\end{equation} 
where: \\ \\
\indent -- the peak of the excitation field is $\mu_{0}h_{0} = 1$ mT,\\
\indent -- the temporal shape of the pulse is $g(t)= \textrm{sinc}(\pi f_{c} t)$, with its duration set by $f_{c} = $ 100 GHz [see Fig.~\ref{FigureMicromagneticMethod}(b)]. \\
\indent -- The thickness profile of the stimulus is $q(z)$ [Fig.~\ref{FigureMicromagneticMethod}(e)].\\
\indent -- Its spatial profile is $f(x) = \textrm{sin}(\vec{k}_{x}.\hat{\vec{e}}_{x}~x)$ [Fig.~\ref{FigureMicromagneticMethod}(a)]. \\

The spatial periodicity of the stimulus along $\hat{\vec{e}}_{x}$ is chosen so as to fit an integer number of wavelengths within the slab in the $x$ direction, by setting $\vec{k}_{x} = (2 \pi n/l_{x}) \hat{\vec{e}}_{x}$ with $n \in \mathbb N$. This leads to a wavevector resolution of $\frac 1 2  \textrm{rad}/\mu\textrm{m}$.

The selective excitation of the acoustical mode is depicted in Fig.~\ref{FigureMicromagneticMethod}(f). To excite the sole acoustical modes, the torque from the stimulus must be made uniform across the thickness of the SAF. This is ensured by setting  $q(z)=q_{ac}(z)=1$ and an orientation of the rf field along $\hat{\vec{e}}_{z}$ [see Fig.~\ref{FigureMicromagneticMethod}(f)]. Conversely, to excite the sole optical mode one applies torques that are opposite in the two layers. For an rf field along $\hat{\vec{e}}_{z}$, this condition is ensured by choosing $q(z) =q_{op}(z) = \textrm{sgn}(i-8.5)$ where sgn is the signum function which is $-1$ for the bottom layers ($i\leq 8$) and $+1$ for the top layers ($i>8$). Finally, if one aims to excite both modes, one can just add the two stimuli and use a thickness profile  $q(z) =q_{op}(z) + q_{ac}(z) = \Theta(i-8.5)$, where $\Theta$ is the Heaviside distribution.

The magnetization response to the stimulus is simulated for $t_{s}=5$ ns. This duration appears sufficient to reach the ground state [Fig.~\ref{FigureMicromagneticMethod}(d)]. We perform a Fourier analysis of the magnetization response averaged over one row of cells $\vec{m}_{i}(t)$, aligned in the direction perpendicular to the wavevector; in practice\cite{mouhoub_exchange_2023} we choose the row $\{i=1, ~x=\frac {l_x} {2}\}$. To minimize the spectrum leakage we apodize this averaged moment with a Hann-window function:
\begin{equation}
    A_{\textrm{Hann}}(t) = \left\{
        \begin{array}{ll}
            \cos^2 \frac{\pi (t-t_{s}/2) }{t_{s}} & \textrm{for} \; 0 \leq t \leq t_{s}\\ 
            0 & \textrm{otherwise}
        \end{array}
    \right.
    \label{HannFunction}
\end{equation}
We then make a Fourier Transform of the apodized form of $\vec{m}_{i=1}(t)$:
\begin{equation}
        \tilde{\vec{m}}_{i = 1}(f)= \mathcal{F} \left( A_{\textrm{Hann}}(t) \vec{m}_{ i = 1}(t) \right)
        \label{fourier}
\end{equation}
where "  $\,\tilde{ }\,$  " recalls the complex-valued nature of frequency-domain magnetization. The SWs frequencies are finally identified by the maxima of the power spectra of the $x$ component of magnetization $\, \tilde{\vec{m}}_{i = 1}(f) \, $:
\begin{equation}
        \max\left(||\tilde{\vec{m}}^{x}_{i = 1}(f)||^2\right). 
        \label{psd}
\end{equation}

One representative result of this procedure (Eq.~\ref{simulus1} to \ref{psd}) is illustrated in Fig.~\ref{FigureMicromagneticMethod}(e). When the thickness-resolved stimulus $q(z)$ is meant to excite both optical and acoustical SWs, two eigenfrequencies are expected if the situation is reciprocal. However in a non-reciprocal situation like ours, 4 eigenfrequencies are generally obtained. This arises because the spatial stimulus $f(x) = \textrm{sin}(\vec{k}_{x}.\hat{\vec{e}}_{x}~x)$ is a standing wave composed of the sum of waves of wavevectors $\vec{k}_{x}$ and $-\vec{k}_{x}$ [~notice that  $f(x) \propto e^{(i \vec{k}_{x}.\hat{\vec{e}}_{x}~ x)}- e^{-(i \vec{k}_{x}.\hat{\vec{e}}_{x}~ x)}$~]. As a result, even if exciting only the acoustical branch [Fig.~\ref{FigureMicromagneticMethod}(f)] or only the optical branch (not shown), one is still left with a pair of eigenfrequencies $\{f_0, f_0'\}$ that must be attributed to either $\{+\vec{k}_{x}, -\vec{k}_{x}\}$ or $\{-\vec{k}_{x}, +\vec{k}_{x}\}$. 
%Deciding which of these two assignments is the correct one requires to implement the upgraded procedure defined so forth.
\subsubsection{Sign-of-k-resolved dispersion relations} \label{signksection}
\begin{figure}[t]
\includegraphics[scale=0.2]{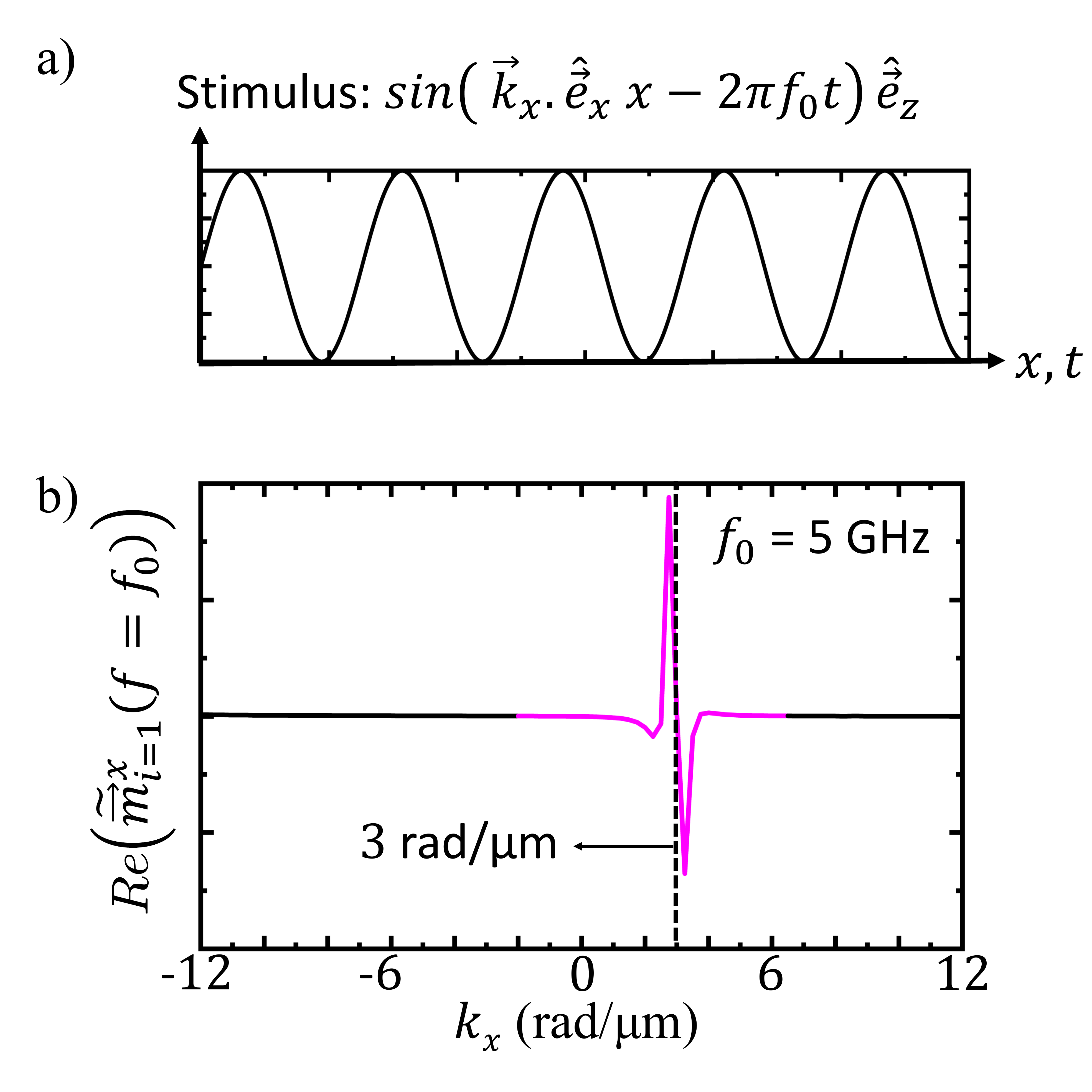}
\caption{Method used to construct the sign-of-k-resolved dispersion relation, as described in section \ref{signksection}. a) Travelling wave like stimulus used to excite.
b) Real part of the dynamical magnetic response of the system at $f_{0} = 5$ GHz versus wavevector according to Eq.~\ref{kResolvedSusceptibility}.}
\label{FigSgnK}
\end{figure}

Getting sign-of-k-resolved dispersion relations requires to apply a stimulus that exclusively excites either the $+\vec{k}_{x}$ or the  $-\vec{k}_{x}$ spin wave [Fig.~\ref{FigSgnK}(a)]. We thus replace the previously used broadband stimuli of Eq.~\ref{simulus1} by \textit{traveling wave} stimuli at the previously identified $f_0$. The new stimuli have the form [see Fig.~\ref{FigSgnK}(a)]:
\begin{equation}
	\vec{H}^{rf} (x,t,z) = h_{0} \, q(z) \, \sin(\vec{k}_{x}.\hat{\vec{e}}_{x}~ x - 2 \pi f_{0} t) \, \hat{\vec{e}}_{z}
	\label{rf-like}
\end{equation}
To ensure that at least one of these stimuli is resonant with $f_0$, we investigate now \textit{signed} wavevectors, i.e. spanning over the bipolar interval: 
\begin{equation}\vec{k}_{x} = (2 \pi n/l_{x}) \hat{\vec{e}}_{x} \textrm{~~with~~} n \in \mathbb Z. \end{equation} 
A much smaller value of $\mu_{0}h_{0}=1~\mu$T is chosen for the (continuously applied) stimulating field to remain in the linear domain.\\
To determine whether a given $\{\vec k_x, f_0\}$ combination is resonant, the relevant information is the susceptibility i.e. the amplitude of the magnetization response once the steady state regime is reached. As before, we focus on one particular row of cells [~e.g. $\{i=1, ~x=\frac {l_x} {2}\}$~] on which we average the magnetizations. We apodize the time response using Eq.~\ref{HannFunction} to minimize the spectral leakage and in addition, to cancel the spectral contribution of the transient response of the magnetization. This requires to calculate the magnetization response for a longer duration, e.g. $t_s=10$ ns.

Since the power of the stimulus is fixed, the resonant or non-resonant character of the dynamical magnetic response versus  $\{ \vec k_x, f_0 \}$ can be evaluated by simply plotting
\begin{equation}
\mathcal {R}e \Big( \mathcal{F} \big( A_{\textrm{Hann}}(t). \vec{m}^{x}_{ i = 1}(t) \big)  \Big) (f=f_0) \label{kResolvedSusceptibility}
\end{equation}
at a given frequency $f_0$ versus $\vec k_x.  \hat{\vec{e}}_{x}$. 
%Sketch of the response for the top view of magnetization is given in Fig.~\ref{FigSgnK}(b,c) demonstrating the response of backward and forward propagation of SWs. Black filled vertical blocks shows the strength of the system's response for a frequency $f_0=5$ GHz [that was identified formerly in Fig.~\ref{FigureMicromagneticMethod}(e) to be a SW frequency] and positive wavevector.
The system's response for a frequency $f_0=5$ GHz versus the wavevector is shown in Fig.~\ref{FigSgnK}(b). From this plot, we can conclude that the specific frequency $f_0$ corresponds to $k_{x}=+3$ rad/$\mu$m and not to $k_{x}=-3$ rad/$\mu$m [see Fig.~\ref{FigSgnK}(b)].
By performing this analysis for each SW frequency that was identified from the broadband stimulus, we can assign each eigenfrequency to a wavevector of the correct sign, and thus we can construct the dispersion relation even in non-reciprocal cases.

\subsection{Dispersion Relations}
\begin{figure}
\includegraphics[scale=0.22]{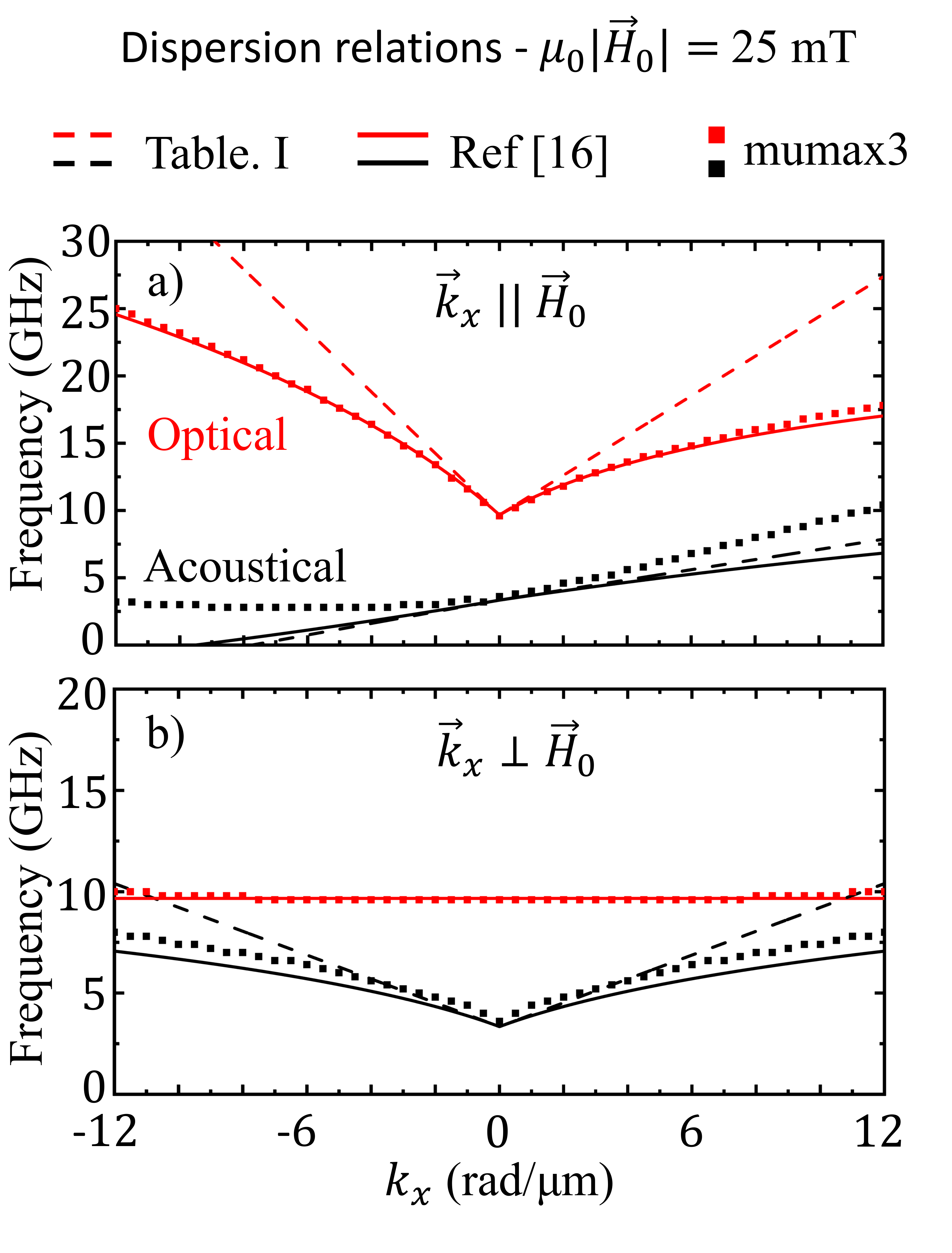}
\caption{Comparison between our model (Table. ~\ref{TableFormulas}, dashed lines), Ishibashi's analytical model\cite{ishibashi_switchable_2020} (continuous lines) and mumax3 simulation (squared dots) adapted with our parameters of Si / SiO$_{x}$ / SAF system for an applied DC field 25 mT. a) Spin waves with $\vec{k}_{x} \parallel \vec{H}_{0}$ (non-reciprocal situation). b) Spin waves with $\vec{k}_{x} \perp \vec{H}_{0}$ (reciprocal situation). For the optical mode, the dashed and continuous lines are superimposed.}
\label{FigDispersionRelation}
\end{figure}

Fig. \ref{FigDispersionRelation} shows the micromagnetic dispersion relations for $\Vec{k}_{x} \parallel \Vec{H}_0$ and $\Vec{k}_{x}\perp \Vec{H}_0$ at $\mu_{0} |\Vec{H}_{0}| = 25$ mT.\\
\indent When $\Vec{k}_{x} \parallel \Vec{H}_0$, a strongly non-reciprocal behaviour \textcolor{black}{due to layer-to-layer dipolar interactions\cite{gallardo_reconfigurable_2019}} is observed for both modes, [see Fig.~\ref{FigDispersionRelation}(a)]. The dispersion relation of the optical mode has the shape of a tilted $\vee$. The dispersion relation of the acoustical mode still exhibits its quite unique feature: it is \textit{essentially linear} at low wavevectors. 
Micromagnetic calculations confirm that the amplitude of the non-reciprocity is very strong:\\
For instance for $|\vec{k}| = 2$ and 12 rad/$\mu$m, it predicts respectively $\delta f_{ac} =  1.6$ and 7.2 GHz, and exactly opposite values $\delta f_{op} = -1.6$ and $-7.2$ GHz. These equalities confirm the analytical prediction of Eq. 3.\\
\indent Conversely, when $\Vec{k}_{x} \perp \Vec{H}_0$, a perfectly reciprocal behavior is observed [Fig.~\ref{FigDispersionRelation}(b)]. An additional striking feature that is specific to SAFs is that the optical mode has a very flat dispersion relation (i.e. $V_{g}^\perp \approx 0$ km/s, $\forall k$). \\
\indent To discuss these micromagnetic dispersion relations, we have superimposed the predictions of two analytical 2-macrospin models based on dynamic matrix theory\cite{
nortemann_microscopic_1993, zhang_angular_1994, grimsditch_magnetic_2004, giovannini_spin_2004}: our formalism (Eq.~\ref{FMRfrequencies} and Table \ref{TableFormulas}) and that of Ishibashi et al. \cite{ishibashi_switchable_2020}. These analytical models do not take into account the limited intralayer exchange stiffness. 

\section{Discussion} %%%%%%%
In this section, we compare the approximate models to the micromagnetic calculations. The results for $\vec{k}_{x} \parallel \vec{H}_{0}$ and $\vec{k}_{x} \perp \vec{H}_{0}$ shown in Fig.~\ref{FigDispersionRelation}(a,b) are completed by additional field magnitudes given in Supplemental Material [Fig.~S1]. We also provide angle-resolved dispersion relations  for  $\mu_{0} |\Vec{H}_{0}| = 25$ mT  in  Supplemental Material [Fig.~S2]. The fields span from 10 mT, for which the scissors state is close to antiparallel with magnetizations that are quite uniform across the two CoFeB layers, up to 75 mT, i.e. close to $\mu_0H_J=78~\textrm{mT}$ where the SAF is close to the parallel state but with a very significant gradient of the magnetization orientation within each CoFeB layer\cite{mouhoub_exchange_2023}.
Several points arise from these comparisons:\\
\indent(i) Uniform resonance ($k$ = 0). At low fields Eq.~\ref{FMRfrequencies} correctly accounts for the precession frequencies. As the applied field increases, the magnetizations of each layer develop an increasing gradient within thickness, and Eq.~\ref{FMRfrequencies} progressively fails to account for the two precession frequencies [see Fig.~S1]. This failure of the 2-macrospin model for nearly saturated thick SAF was already noticed and discussed in ref. [\onlinecite{mouhoub_exchange_2023}]. An unavoidable consequence of this failure of the 2-macrospin model at large fields is that the group velocities of Table~\ref{TableFormulas} and of Ishibashi et al.\cite{ishibashi_switchable_2020}, which are both calculated assuming a 2-macrospin ground state, are increasingly failing when approaching saturation.\\
\indent (ii) Concerning the optical mode, a thorough examination of Fig.~\ref{FigDispersionRelation} and 
Fig.~S1 indicates that the three models agree for low and moderate applied fields, i.e. $\mu_0 H_{0} \leq 50 ~\textrm{mT}$ [see Fig.~S1]. The linear expansions of Table \ref{TableFormulas} are usable for wavevectors $ k \in [-12, 12]$ rad/$\mu$m when $\vec{k}_{x} \perp \vec{H}_{0}$: the dispersion relation $f_{op}(k)$ is then very flat and one can hardly perceive some curvature of at the largest wavevectors [see Fig.~\ref{FigDispersionRelation}(b)]. For $\vec{k}_{x} \parallel \vec{H}_{0}$, the linear expansions of Table \ref{TableFormulas} are usable only up to $ k \in [-3, 3]$ rad/$\mu$m. Above 3 rad/$\mu$m, Ishibashi's model performs better and almost agrees with the result of micromagnetic calculations [see Fig.~\ref{FigDispersionRelation}(a)]. This agreement indicates that the intralayer exchange stiffness --neglected in the approximate models-- has little influence on the optical dispersion relation at least up to 12 rad/$\mu$m.\\
\indent (iii) Concerning the acoustical mode, for $\vec{k}_{x} \perp \vec{H}_{0}$, the approximate models perform well at low and moderate applied fields. At fields $H_{0} \leq 15$ mT [see Fig.~S1(b),(f)], the predictions of Table I are closer to full micromagnetic calculations than those of ref.~[\onlinecite{ishibashi_switchable_2020}]. At these fields the branches of the $\vee$-shaped dispersion relation are remarkably close to linear. The branches acquire progressively a curvature for $\mu_0 H_{0} \geq 25$ mT [see Fig.~\ref{FigDispersionRelation}(a)], and then Ishibashi's model start performing better than the expressions of Table I. \\
The case of $\vec{k}_{x} \parallel \vec{H}_{0}$ is far less satisfactory [Fig.~\ref{FigDispersionRelation}(a)]. The approximate models perform reasonably well but only at very low $k \in [-3, 3]$ rad/$\mu$m, and provided that the applied field is low. 
For instance at a field of 25 mT, a linear fit through positive wavevector yields $V_{g}=2.95$ km/s for micromagnetic calculations [Fig.~\ref{FigDispersionRelation}(a)] to be compared with $V_{g} = 2.57$ km/s from Table I and $V_{g}=2.4$ km/s from Ishibashi's analytical model. Besides, these three velocities are higher than the experimental results which were $V_{g} (k_{x}>0) = 1.25 $ km/s for Si / SiO$_{x}$ / SAF system studied by BLS [see Fig.~\ref{FigureBLS}(c)], and $V_{g}(k_{x}>0) = 1.9 $ km/s for LiNbO$_{3}$ / SAF system by propagating spin wave spectroscopy [see Fig.~\ref{FigToggling}(d,e,f,g)]. \\
\indent There is thus a discrepancy for $\vec{k}_{x} \parallel \vec{H}_{0}$ both among the models and with the experimental results. We believe that this is an indication that the texture of the ground state --it being different from a 2-macrospin scissors-- has a major impact on the dispersion relation of the acoustical mode for this field orientation. As mentioned in section \ref{PeculiaritiesDispersionRelations}, the slope of the dispersion relation of the acoustical mode is set by the scaling factor $\frac{1}{2} \gamma_0 M_s t_{\textrm{mag}}$: it is proportional to the thickness $t_{\textrm{mag}}$ of each layer; if a gradient of the magnetization develops within each layer, the relevant thickness to be considered becomes ill-defined. This is also probably the case in experimental samples where the exchange stiffness is potentially non-uniform near the Ru barrier because of composition gradients\cite{seeger_inducing_2023}. This has a substantial impact on the acoustical dispersion relation for $\vec{k}_{x} \parallel \vec{H}_{0}$, which remains to be fully expressed analytically. 

\section{Conclusions}
In summary, we have combined experiments, analytical modeling and numerical simulations to study the spin waves in symmetric synthetic antiferromagnets in the scissors state. The non-reciprocity comes from the layer-to-layer dipolar interactions, which is the only contribution . Whatever the wavevector orientation, the frequency non-reciprocities of the acoustical and optical spin waves coincide in magnitude but have opposite signs. Our modeling explains why the frequency non-reciprocity can reach unusually large magnitudes, especially when the static field is oriented close to the spinwave wavevector [Fig.~S2]. 
In this case, the acoustical spin waves exhibit a unique dispersion relation: it is close to linear around $k=0$. As a result, reversing the sign of the SW wavevector does not reverse of the direction of the group velocity, such that the energy flow associated to a wavepacket of acoustical spin waves is truly unidirectional: these wavepackets can carry energy in one direction but not the opposite.
\textcolor{black}{We have developed a simple analytical model to approximate group velocities and evidenced switchable unidirectionality of spin waves in synthetic antiferromagnets, a feature that opens potential applications.} 

\section*{Acknowledments}
The authors acknowledge the French National Research Agency (ANR) under contract N$^{o}$ ANR-20-CE24-0025 (MAXSAW). This work was supported by a public grant overseen by the French National Research Agency (ANR) as part of the “Investissements d'Avenir” program (Labex NanoSaclay, reference: ANR-10-LABX-0035, project SPICY). This work was done within the C2N micro nanotechnologies platforms and partly supported by the RENATECH network and the General Council of Essonne.

\subsection*{Appendix: Sample details} %%%
Piezoelectric Y-cut LiNbO$_{3}$ substrates were used for the fabrication of devices meant for electrical measurements. A strong interlayer exchange coupling $J = -1.7$ mJ/m$^{2}$ is obtained when using these substrates \cite{mouhoub_exchange_2023, seeger_inducing_2023}, leading to an interlayer exchange field $\mu_{0} H_{J} = -\frac{2J}{ M_{s} t_{\textrm{mag}}}$ which amounts to 148 mT. The relatively low exchange stiffness $A_{\textnormal{ex}} = 16 \; \textnormal{pJ/m}$ within this SAF allows for gradients of magnetization orientation within the thicknesses of the CoFeB layers \cite{devolder_measuring_2022, mouhoub_exchange_2023}, which increases the saturation field up to $\mu_{0}H_\textrm{sat} = 251$ mT. There is a small (unwelcome) uniaxial anisotropy field, with anisotropy field being $\mu_{0}H_{k} = 3$ mT and spin-flop field $\mu_{0}H_{\textrm{SF}} = 19$ mT.

Si / SiO$_{x}$ substrates were used instead of the piezoelectric LiNbO$_3$ substrates when preparing samples meant for Brillouin Light scattering. Using Si / SiO$_{x}$ substrates ensures that the BLS signals can be unambiguously attributed to spin waves and not to the surface or bulk acoustical waves potentially present in the LiNbO$_3$ substrates. On these Si / SiO$_{x}$ substrates, a thermal treatment can be applied to suppress any anisotropy\cite{seeger_inducing_2023}. It was effectively reduced to only $\mu_{0}H_{k} = 0.8$ mT. The thermal treatment also reduces the interlayer exchange energy down to $J = -0.9$ mJ/m$^{2}$ (i.e. $\mu_{0} H_{J} = 78$ mT)  while it increases the exchange stiffness $A_{\textnormal{ex}} = 28.3 \; \textnormal{pJ/m}$. Spin-flop field amounts to  $\mu_{0}H_{\textrm{SF}} = 8$ mT.
As a result of the strong intralayer exchange, this SAF behaves more like a 2-macrospin system and its saturation field $\mu_{0}H_\textrm{sat} = 100$ mT is thus close to its $\mu_{0} H_{J}$. \newpage

\widetext
\begin{center}
	\textbf{\large Supplemental Material: Unidirectionality of spin waves in Synthetic Antiferromagnets} \newline
	
	F. Millo,$^{1 \textrm{, a)}}$ J.-P. Adam,$^{1}$ C. Chappert,$^{1}$ J.-V. Kim,$^{1}$ A. Mouhoub,$^{1}$ A. Solignac,$^{2}$ and T. Devolder$^{1}$\newline
 	
	$^{1)}$ \textit{Université Paris-Saclay, CNRS, C2N, 91120 Palaiseau, France} \\
	$^{2)}$\textit{SPEC, CEA, CNRS, Université Paris-Saclay, 91191 Gif-sur-Yvette, France}

\end{center}
\setcounter{figure}{0}
\renewcommand{\figurename}{FIG.}
\renewcommand{\thefigure}{S\arabic{figure}}
\setcounter{equation}{0}
\setcounter{figure}{0}
\setcounter{table}{0}
\setcounter{page}{1}
\begin{figure}[!htb]
	\includegraphics[scale=0.13]{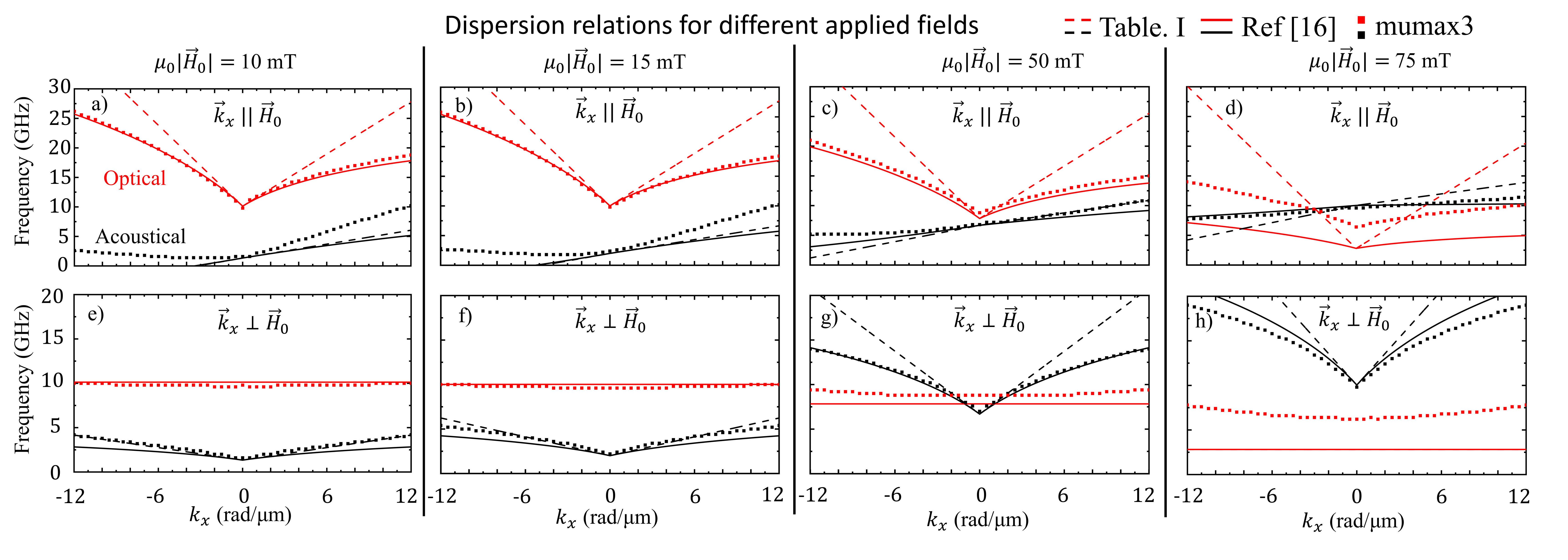}
	\caption{Comparison between our model (Table. I, dashed lines), Ishibashi's analytical model$^\textrm{16}$ (continuous lines) and mumax3 simulation (squared dots) adapted with parameters of Si / SiO$_{x}$ / SAF system for different applied DC fields (column representation). a-d) $\vec{k}_{x} || \vec{H}_{0}$ is the non-reciprocal situation. e-h) $\vec{k}_{x} \perp \vec{H}_{0}$ is the reciprocal situation. For optical mode, the dashed and continuous lines are superimposed.}
	\label{FigSuppS1}
\end{figure}

\begin{figure}[!htb]
	\includegraphics[scale=0.13]{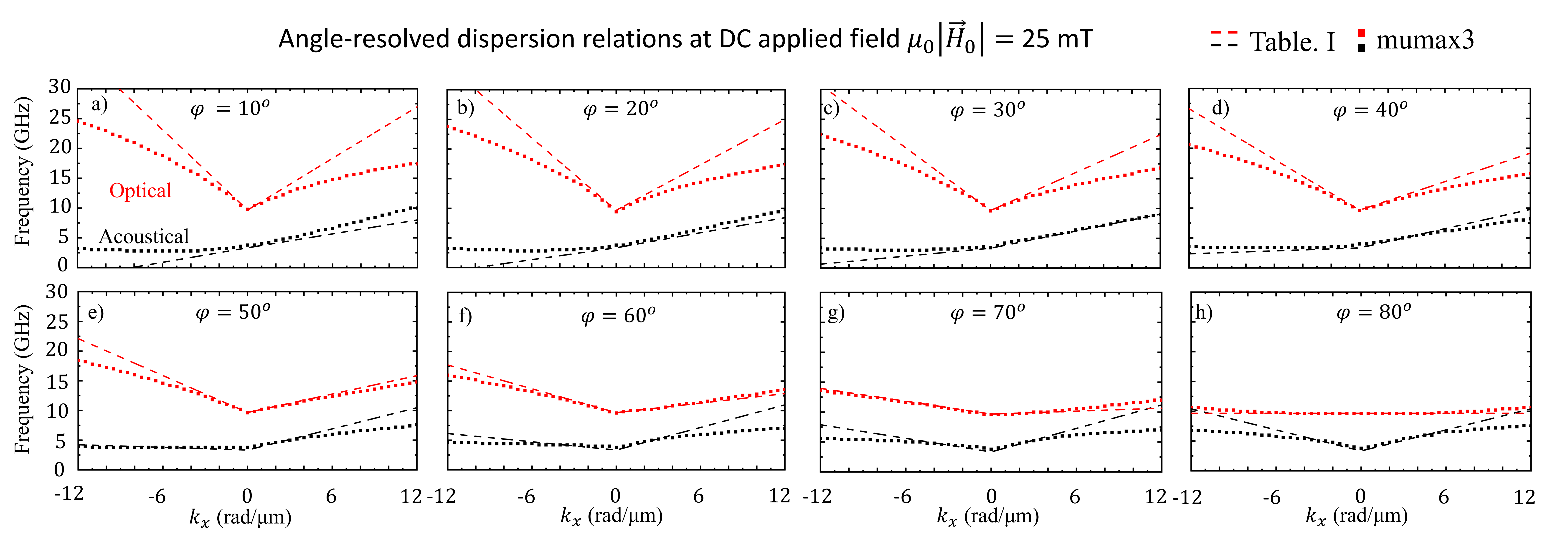}
	\caption{ Dispersion relations for different angles $\varphi$ (defined between $\vec{k}_{x}$ and $\vec{H}_{0}$) at DC applied field $\mu_{0}|\vec{H}_{0}|=25$ mT.}
	\label{FigSuppS2}
\end{figure}

\end{document}